\documentclass[prd,amsmath,superscriptaddress,twocolumn]{revtex4}

\usepackage{hyperref}
\usepackage{ifthen}
\usepackage{amsmath}
\usepackage{amssymb}
\usepackage{graphics}
\usepackage{color}

\definecolor{darkgreen}{cmyk}{0.85,0.2,1.00,0.2}
\definecolor{purple}{cmyk}{0.5,1.0,0,0}

\begin{document}

\title{Constraints on $f(R)$ gravity from probing the large-scale structure}

\author{Lucas Lombriser}
\affiliation{Institute for Theoretical Physics, University of Z\"{u}rich, Winterthurerstrasse 190, CH-8057 Z\"{u}rich, Switzerland}
%\affiliation{Department of Physics, ETH Z\"{u}rich, CH-8093 Z\"{u}rich, Switzerland}
\author{An\v{z}e Slosar}
\affiliation{Brookhaven National Laboratory, Physics Department, Upton NY 11973, USA}
\author{Uro\v{s} Seljak}
\affiliation{Institute for Theoretical Physics, University of Z\"{u}rich, Winterthurerstrasse 190, CH-8057 Z\"{u}rich, Switzerland}
\affiliation{Physics and Astronomy Department, University of California, and Lawrence Berkeley National Laboratory, Berkeley, California 94720, USA}
\affiliation{Ewha University, Seoul 120-750, Korea}
\author{Wayne Hu}
\affiliation{Kavli Institute for Cosmological Physics, Department of Astronomy and Astrophysics, Enrico Fermi Institute, University of Chicago, Chicago, Illinois 60637, USA}

\date{\today}

\begin{abstract}
We study cosmological constraints on metric $f(R)$ gravity models that are designed to reproduce the $\Lambda$CDM expansion history with modifications to gravity described by a supplementary cosmological freedom, the Compton wavelength parameter $B_0$. We conduct a Markov chain Monte Carlo analysis on the parameter space, utilizing the geometrical constraints from supernovae distances, the baryon acoustic oscillation distances, and the Hubble constant, along with all of the cosmic microwave background data, including the largest scales, its correlation with galaxies, and a probe of the relation between weak gravitational lensing and galaxy flows. The strongest constraints, however, are obtained through the inclusion of data from cluster abundance. Using all of the data, we infer a bound of $B_0<1.1\times10^{-3}$ at the 95\% C.L.
\end{abstract}

\maketitle

\section{Introduction}

Cosmic acceleration can either be explained by introducing large amounts of dark energy or considering modifications to gravity such as the addition of a suitable function $f(R)$ of the Ricci scalar to the Einstein-Hilbert action~\cite{carroll:03, nojiri:03, capozziello:03}. In fact, one may interpret the cosmological constant as being of this kind rather than attributing it to vacuum energy. It has been argued that a valid $f(R)$ model should closely match the $\Lambda$CDM expansion history~\cite{hu:07a, brax:08}. We specialize our considerations to functions $f(R)$ that exactly reproduce this background and parametrize the class of solutions in terms of its Compton wavelength parameter $B_0$~\cite{song:06}. Such $f(R)$ modifications affect gravity at solar-system scales, which are well tested and impose stringent constraints on deviations from general relativity. However, the chameleon effect~\cite{khoury:03, navarro:06, faulkner:06} provides a mechanism that allows certain $f(R)$ models to evade solar-system tests (e.g.,~\cite{hu:07a}). The transition required to interpolate between the low curvature of the large-scale structure and the high curvature of the galactic halo sets the strongest bound on the cosmological field ($B_0\lesssim10^{-5}$~\cite{hu:07a, smith:09t}).

Independently, strong constraints can also be inferred from the large-scale structure alone.
The enhanced growth of structure observed in $f(R)$ gravity models manifests itself on the largest scales of the cosmic microwave background (CMB) temperature anisotropy power spectrum~\cite{song:06}, where consistency with CMB data places an upper bound on $B_0$ of order unity~\cite{song:07}. Cross correlations of the CMB temperature field with foreground galaxies serve as another interesting test of $f(R)$ gravity models~\cite{song:06, song:07}, tightening the constraint on the Compton wavelength parameter by an order of magnitude (e.g.,~\cite{giannantonio:09}). However, the currently strongest constraints on $f(R)$ gravity from large scale structures are inferred from the analysis of the abundance of low-redshift X-ray clusters, yielding an improvement over the CMB constraints on the free field amplitude of the Hu-Sawicki~\cite{hu:07a} ($n=1$) model of nearly four orders of magnitude~\cite{schmidt:09}.

In this paper, we perform an independent analysis of $f(R)$ gravity constraints, focusing on cosmological data only.
%For the first time combine all of the observables discussed above within the designer model of $f(R)$ gravity.
Our analysis differs from previous studies partially in terms of the theoretical model, the parametrical approach, or data sets implemented. We compare and discuss our results to the constraints of previous analyses. Here, we conduct a Markov chain Monte Carlo (MCMC) study of metric $f(R)$ gravity models that are designed to reproduce the $\Lambda$CDM expansion history using data from CMB anisotropies, supernovae distances, the baryon acoustic oscillation (BAO) distances, and the Hubble constant. For observables in the linear regime, we adopt the parametrized post-Friedmann (PPF) framework~\cite{hu:07b, hu:08} and its implementation into a standard Einstein-Boltzmann linear theory solver~\cite{fang:08} for the theoretical predictions. This framework allows us to include information from the near horizon scales. We also utilize information from the cross correlation between high-redshift galaxies and the CMB through the integrated Sachs-Wolfe (ISW) effect.
%, which has been proposed as an interesting test of $f(R)$ gravity~\cite{song:06, song:07}.
We further use a probe of the relation between weak gravitational lensing and galaxy flows, as well as data from the abundance of clusters that are identified by overdensities of bright, uniformly red galaxies.
Latter yields the tightest constraints on the cosmological parameters, particularly on $B_0$.
%(cf.~\cite{schmidt:09, rapetti:09}).
We compare these constraints to the results of~\cite{schmidt:09} derived for the Hu-Sawicki model.

In \textsection{\ref{sec:theory}}, we review metric $f(R)$ gravity theory. We present the results of our MCMC study in \textsection{\ref{sec:constraints}} and discuss them in \textsection{\ref{sec:discussion}}. Finally, the PPF formalism for $f(R)$ gravity~\cite{hu:08} and details about the modifications to the {\sc iswwll} code~\cite{ho:08, hirata:08} used for the galaxy-ISW (gISW) cross correlation observations are specified in the Appendix.

\section{$f(R)$ Gravity} \label{sec:theory}

In $f(R)$ gravity, the Einstein-Hilbert action is supplemented by a free function of the Ricci scalar $R$,
\begin{equation}
S = \frac{1}{2\kappa^2}\int d^4x \sqrt{-g}[R+f(R)] + \int d^4x \sqrt{-g}\mathcal{L}_{\rm m}.
\end{equation}
Here, $\kappa^2\equiv 8\pi G$ and $\mathcal{L}_{\rm m}$ is the matter Lagrangian, where we have set $c=1$. Variation with respect to the metric $g_{\alpha\beta}$ yields the modified Einstein equation for metric $f(R)$ gravity,
\begin{equation}
G_{\mu\nu} + f_R R_{\mu\nu} - \left( \frac{f}{2} - \Box f_R \right) g_{\mu\nu} - \nabla_{\mu}\nabla_{\nu} f_R = \kappa^2 T_{\mu\nu},
\label{eq:einstein}
\end{equation}
where subscripts of $R$ indicate differentiation with respect to the Ricci scalar and the connection is of Levi-Civita type.
The modified Friedmann equation is derived in the usual way, i.e., taking the time-time component of Eq.~(\ref{eq:einstein}),
\begin{equation}
H^2 - f_R(HH'+H^2) + \frac{f}{6} + H^2 f_{RR} R' = \frac{\kappa^2}{3}\rho,
\label{eq:friedmann}
\end{equation}
where here and throughout the paper primes denote derivatives with respect to $\ln a$.

\subsubsection{Designer model}

The function $f(R)$ can be constructed in a way that it recovers any background history. Given the equation of state of the effective dark energy, $w_{\rm DE}$, one integrates the modified Friedmann equation to obtain the corresponding form of $f(R)$~\cite{song:06, pogosian:07}. It has been pointed out that viable $f(R)$ cosmologies must closely match the $\Lambda$CDM expansion history~\cite{hu:07a, brax:08}. Here, we focus on flat models and consider modifications to gravity that reproduce the $\Lambda$CDM expansion history exactly, i.e., $w_{\rm DE} = -1$, where we neglect contributions from radiation. Note that the restriction to the matter-dominated epoch is well motivated by requiring that the well-tested high-curvature regime reproduces standard phenomenology (e.g.,~\cite{song:06}). The $\Lambda$CDM background is then given by
\begin{equation}
H^2 = \frac{\kappa^2}{3}(\rho_{\rm m} + \rho_{\Lambda}).
\label{eq:}
\end{equation}
Equating it with the matter-dominated Friedmann equation, Eq.~(\ref{eq:friedmann}), yields an inhomogeneous second order differential equation for $f(R)$,
\begin{equation}
f'' - \left[ 1+\frac{H'}{H}+\frac{R''}{R'} \right] f' + \frac{R'}{6H^2} f = -H_0^2(1-\Omega_m)\frac{R'}{6H^2}.
\label{eq:fODE}
\end{equation}
This can be solved numerically with the initial conditions
\begin{eqnarray}
f(\ln a_i) & = & A H_0^2 a_i^p - 6H_0^2\Omega_{\Lambda}, \\
f'(\ln a_i) & = & p A H_0^2 a_i^p,
\end{eqnarray}
where $p=(-7+\sqrt{73})/4$ and $a_i\sim0.01$~\cite{song:06}. $A$ is an initial growing mode amplitude and characterizes a specific solution in the set of functions $f(R)$ that recovers the $\Lambda$CDM background. Note that the amplitude of the decaying mode is set to zero in order to not violate constraints in the high-curvature regime. We follow~\cite{song:06} and parametrize our solutions in terms of the Compton wavelength parameter
\begin{equation}
B = \frac{f_{RR}}{1+f_R} R' \frac{H}{H'}
\end{equation}
evaluated at $B_0\equiv B(\ln a = 0)$ rather than by $A$. In a Taylor expansion of $f(R)$, the first term corresponds to a cosmological constant, while the second is a rescaling of Newton's constant. With its $f_{RR}$ term, this parametrization therefore captures the essence of the modification. Standard gravity is recovered in the case where $B_0=0$. For stability reasons the mass squared of the scalar field $f_R$ must be positive, which implies $B_0\geq0$~\cite{song:06, sawicki:07}.

\section{Cosmological Constraints} \label{sec:constraints}

We use a variety of cosmological data sets to constrain the $f(R)$ theory of gravity. First we use the CMB anisotropy data from the five-year Wilkinson Microwave Anisotropy Probe (WMAP)~\cite{WMAP:08}, the Arcminute Cosmology Bolometer Array Receiver (ACBAR)~\cite{ACBAR:07}, the Cosmic Background Imager (CBI)~\cite{CBI:04}, and the Very Small Array (VSA)~\cite{VSA:03}. Next we employ data from the Supernova Cosmology Project (SCP) Union~\cite{UNION:08} compilation, the measurement of the Hubble constant from the Supernovae and $H_0$ for the Equation of State (SHOES)~\cite{riess:09} program, and the BAO distance measurements of~\cite{BAO:09}. Furthermore, we take gISW cross correlation observations using the {\sc iswwll} code of~\cite{ho:08, hirata:08}, the $E_G$ measurement, probing the relation between weak gravitational lensing and galaxy flows, of~\cite{reyes:10}, as well as cluster abundance constraints from the likelihood code of~\cite{seljak:09} (CA). We quote results with the latter three constraints separately to highlight their impact on the results. For comparison, we also analyze constraints that include the $\sigma_8$ measure of the Chandra Cluster Cosmology Project (CCCP)~\cite{vikhlinin:08}.

In \textsection\ref{sec:predictions} we discuss the predictions for some of these observables for specific values of the Compton wavelength parameter. In \textsection\ref{sec:results} we present the results of a MCMC likelihood analysis, which is conducted with the publicly available {\sc cosmomc}~\cite{lewis:02} package.

\subsection{Cosmological observables} \label{sec:predictions}

In this section we illustrate model predictions of the various cosmological observables we use in the constraints. We choose the parameters of the various models that highlight results from the MCMC analysis.

By construction, at high redshifts, the $f(R)$ modifications become negligible, and so we choose a parametrization that separates high-redshift and low-redshift constraints. Specifically we take 6 high-redshift parameters: the physical baryon and cold dark matter energy density $\Omega_bh^2$ and $\Omega_ch^2$, the ratio of sound horizon to angular diameter distance at recombination multiplied by a factor of 100 $\theta$, the optical depth to reionization $\tau$, the scalar tilt $n_s$, and amplitude $A_s$ at $k_*=0.002~\textrm{Mpc}^{-1}$.

Since restricting to flat universes, at low redshifts, $\Lambda$CDM has no additional parameter, whereas in $f(R)$ gravity, an extra degree of freedom is introduced by the Compton wavelength parameter $B_0$, where $\Lambda$CDM is reproduced in the limit $B_0\rightarrow0$.

We illustrate predictions from the maximum likelihood $\Lambda$CDM model using all of the data (see Table~\ref{tab:res_flat_lcdm_1}). For $f(R)$ gravity, where not otherwise specified, we use these $\Lambda$CDM best-fit parameters and add nonzero values of $B_0$ for comparison.

\subsubsection{Geometrical measures}

The comparison of the magnitudes of high-redshift to low-redshift supernovae yields a relative distance measure. The acoustic peaks in the CMB, the measurement of the local Hubble constant, and the BAO distances additionally provide absolute distance probes which complement the relative distance measure of the supernovae. These probes constrain the background and since our $f(R)$ models are designed to match the $\Lambda$CDM expansion history with vanishing effect in the high-curvature regime, they do not generate any tension between the models. For the Hubble constant, we use the SHOES measurement of
\begin{equation}
H_0=74.2\pm3.6~{\rm km\ s^{-1}\ Mpc^{-1}},
\end{equation}
which employs Cepheid measurements to link the low-redshift supernovae to the distance scale established by the maser galaxy NGC 4258.
Further, we apply the constraint
\begin{equation}
\Omega_{\rm m} = (0.282\pm0.018) \left( \frac{\Omega_{\rm m}h^2}{0.1326} \right)^{0.58}
\end{equation}
from the BAO distance measure of~\cite{BAO:09} that is obtained from
analyzing the clustering of galaxies from the Sloan Digital Sky Survey (SDSS)~\cite{abazajian:09} and the 2-degree Field Galaxy Redshift Survey (2dFGRS)~\cite{cole:05}. This measurement yields the tightest constraint on $\Omega_{\rm m}$ and substantially assists in breaking degeneracies with $B_0$ in the cluster abundance data.

\subsubsection{The cosmic microwave background}

The CMB probes the geometry of the background history as well as the formation of large-scale structure. The latter manifests itself on the largest scales through the ISW effect from the evolution of the gravitational potential. To predict these effects we implement the PPF modifications described in Appendix~\ref{sec:PPF}. The incorporation of the PPF formalism into a standard Einstein-Boltzmann linear theory solver yields an efficient way to obtain predictions of $f(R)$ gravity for the CMB. We utilize the PPF modifications to {\sc camb}~\cite{CAMB:99} implemented in Ref.~\cite{fang:08}, which we configure for $f(R)$ gravity as described by~\cite{hu:08}.

\begin{figure}
\resizebox{\hsize}{!}{\includegraphics{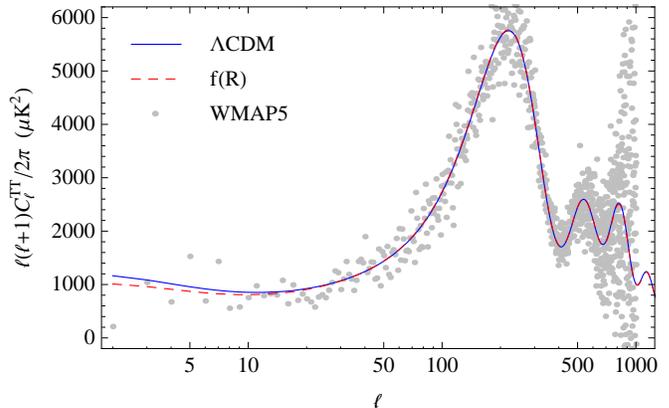}}
\caption{Best-fit CMB temperature anisotropy power spectrum for $\Lambda$CDM and $f(R)$ gravity from using the Union, SHOES, BAO, and CMB data.}
\label{fig:CMB}
\end{figure}

In Fig.~\ref{fig:CMB}, we plot the CMB temperature anisotropy power spectrum with respect to angular multipole $\ell$ for the best-fit $\Lambda$CDM model and the best-fit $f(R)$ gravity model using the Union, SHOES, BAO, and CMB data jointly (see Tables~\ref{tab:res_flat_lcdm_1} and \ref{tab:res_flat_fR_1}). Adding an additional freedom from $f(R)$ gravity only yields an insignificant improvement in the fit.
Relative to $\Lambda$CDM, or equivalently $B_0=0$, the growth of structure is enhanced. The ISW effect at the lowest multipoles is decreasing for $B_0 \lesssim 3/2$~\cite{song:06, song:07}. At $B_0\sim3/2$, the ISW contribution rises again and turns into a relative enhancement over $\Lambda$CDM for $B_0 \gtrsim 3$, ruling out $B_0\gtrsim5$ by WMAP data~\cite{song:06, song:07}.

Due to suppression of $f(R)$ modifications in the high-curvature regime, the CMB acoustic peaks can be utilized as usual to infer constraints on the high-redshift parameters, in particular the physical energy densities of baryonic matter and dark matter as well as the angular diameter distance to recombination.

\subsubsection{Galaxy-ISW cross correlations} \label{gISW}

\begin{figure*}
\resizebox{\hsize}{!}{\includegraphics{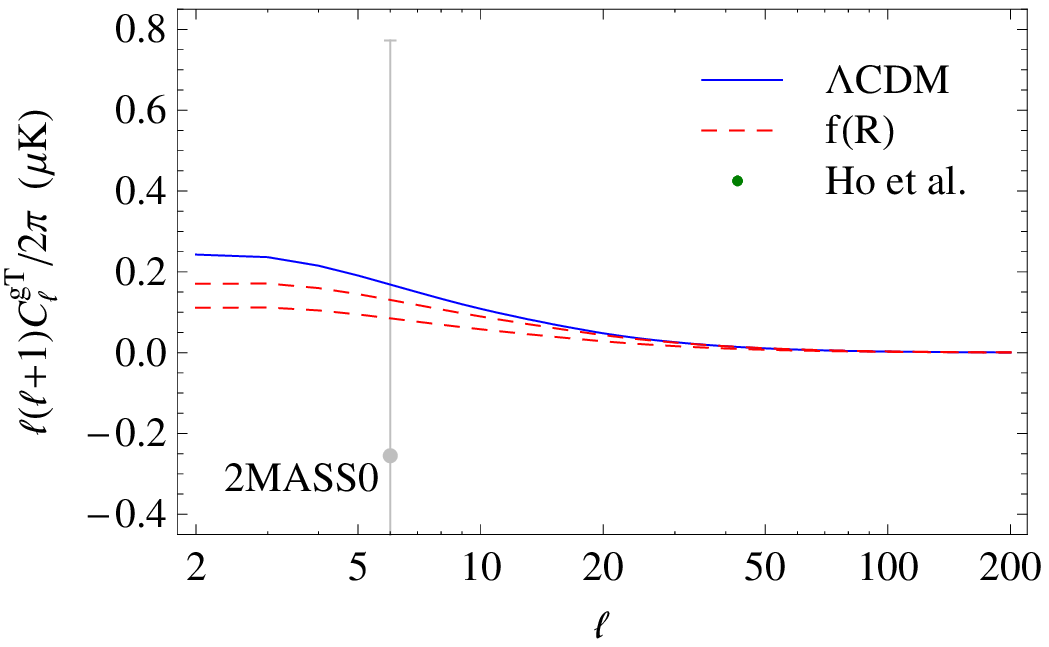}\includegraphics{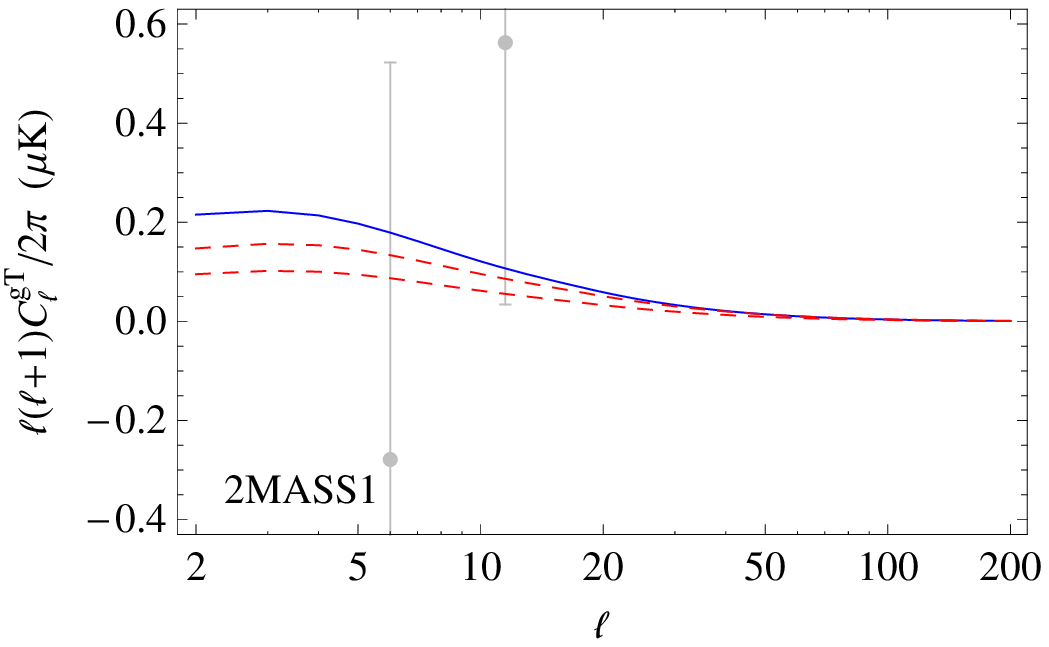}\includegraphics{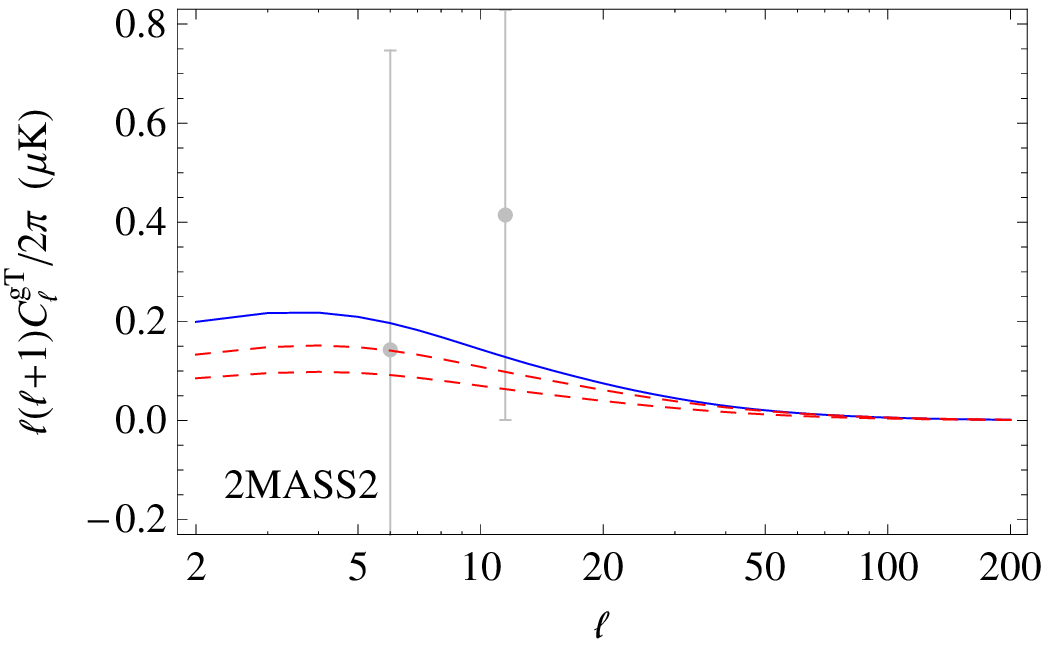}}
\resizebox{\hsize}{!}{\includegraphics{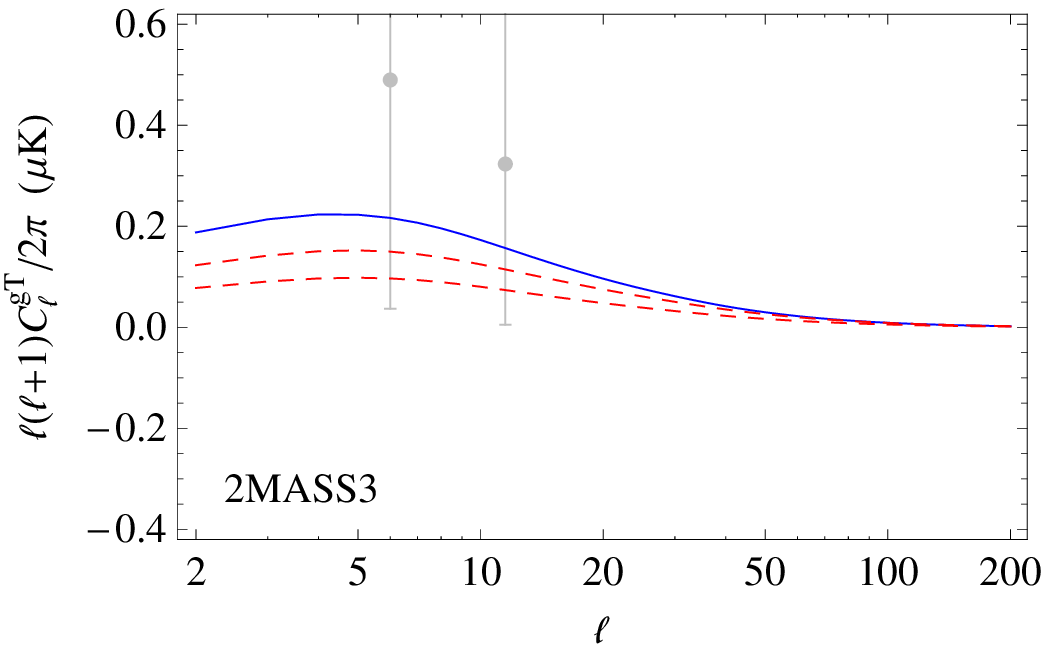}\includegraphics{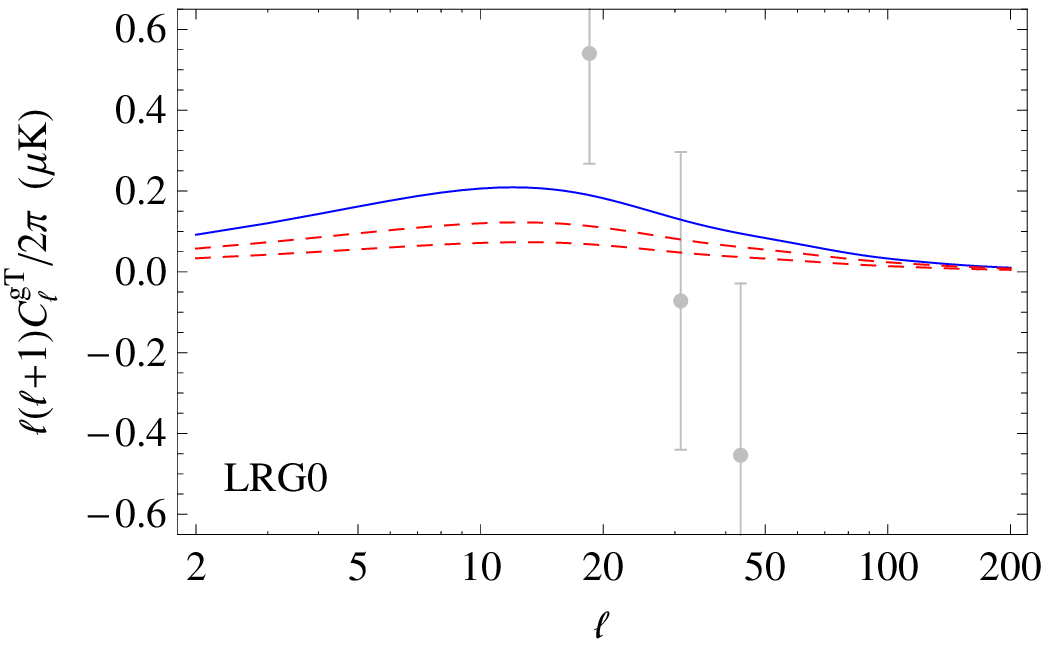}\includegraphics{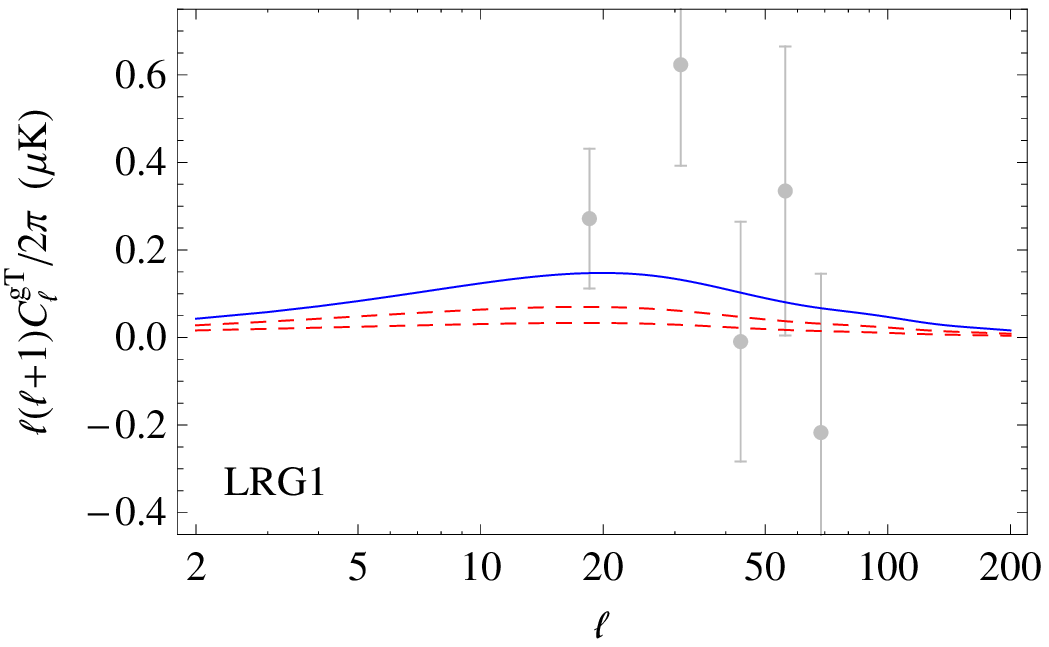}}
\resizebox{\hsize}{!}{\includegraphics{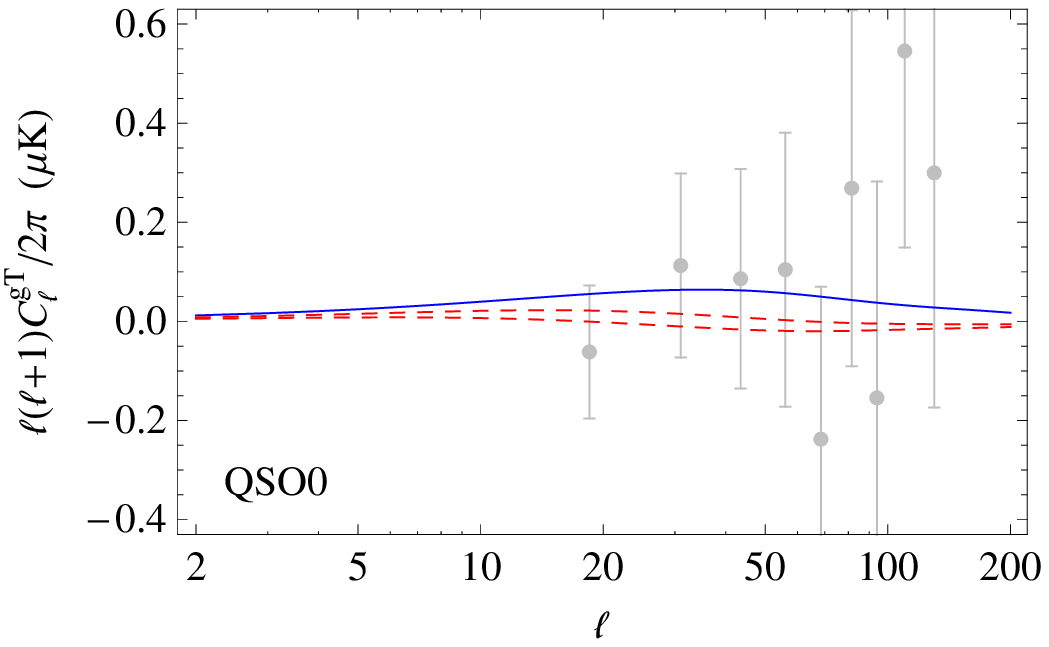}\includegraphics{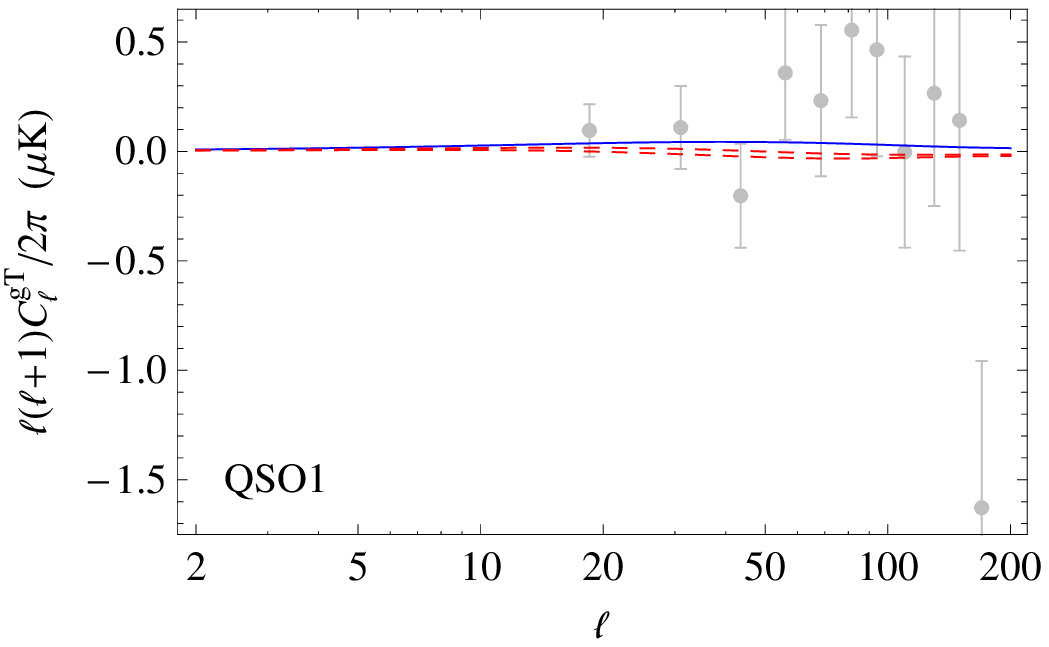}\includegraphics{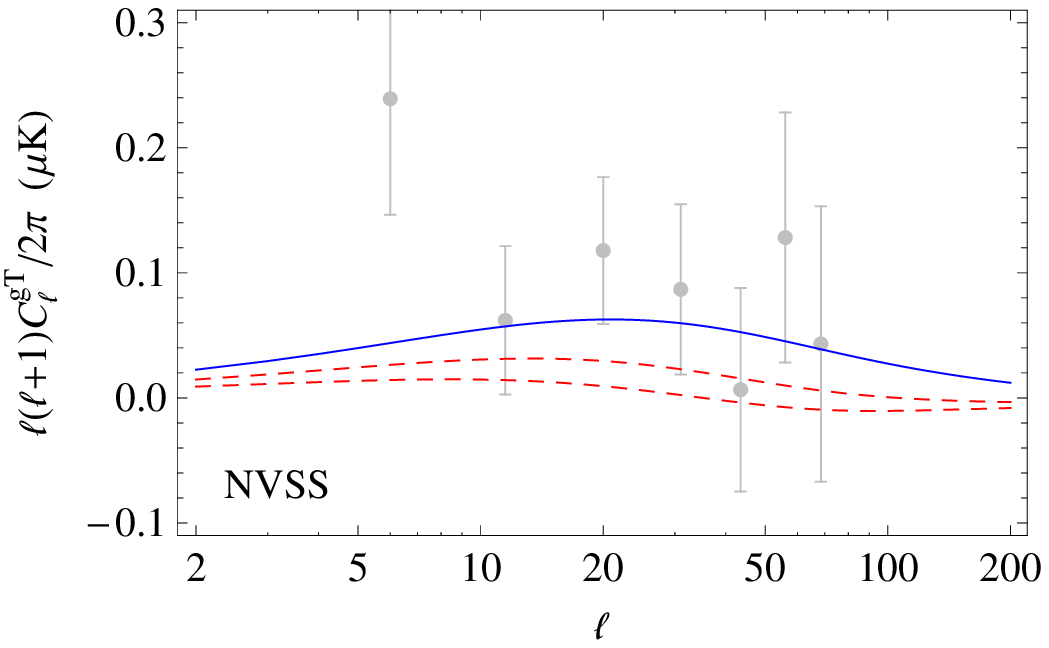}}
\caption{Best-fit $\Lambda$CDM ($B_0=0$) gISW cross correlations (blue solid line) for the different galaxy samples, roughly ordered in increasing effective, bias-weighted, redshift. Adding nonzero values for the Compton wavelength parameter (red dashed lines): $B_0=0.1$ (top) and $B_0=0.5$ (bottom).}
\label{fig:ISW}
\end{figure*}

The correlation between galaxy number densities and the CMB anisotropies can be used to isolate the ISW effect in the CMB. Whereas the ISW effect in the CMB is suppressed for $B_0\lesssim3$ and enhanced for $B_0\gtrsim3$ with respect to $B_0=0$, the gISW cross correlation is suppressed for all $B_0>0$ and leads eventually to anti-correlations (see Fig.~\ref{fig:ISW}). This makes the gISW cross correlation interesting for improving constraints (see, e.g.,~\cite{song:06, song:07}). As was shown in~\cite{song:07}, the absence of negative correlations between the CMB and an assortment of galaxy surveys infer a boundary of $B_0\lesssim1$. This constraint can be improved by a more rigorous analysis of gISW cross correlations as was performed by, e.g.,~\cite{giannantonio:09}. This study implements a parametrization for the modifications induced by $f(R)$ gravity that is based on the introduction of an effective scalar degree of freedom in the Einstein-Hilbert action~\cite{bertschinger:08, zhao:08} and uses gISW cross correlation data from~\cite{giannantonio:08}. Here, we conduct an independent analysis, utilizing the PPF framework and for the calculation of the gISW likelihood, the {\sc iswwll} code of~\cite{ho:08, hirata:08}, which has turned out to be very useful for constraining infrared modifications of gravity~\cite{lombriser:09, bean:10, daniel:10b, lombriser:11}.

We evaluate the gISW cross correlation in the Limber and quasistatic approximation, as it is done in the {\sc iswwll} code used for the data analysis. The gISW cross correlation in this approximation reads
\begin{eqnarray}
C_{\ell}^{g_j T} & \simeq & \frac{3\Omega_m H_0^2 T_{\rm CMB}}{(l+1/2)^2} \int dz f_j(z) H(z) \nonumber\\
& & \times \left. \left[ D(a,k) \frac{d}{dz} G(a,k) P(k) \right] \right|_{k=\frac{l+1/2}{\chi(z)}}
\label{eq:gISW}
\end{eqnarray}
with $z=1/a-1$ denoting redshift.
Here, $D(a,k)$ is the linear growth rate in the quasistatic regime defined by $\Delta_{\rm m}(a,k) = \Delta_{\rm m}(1,k) D(a,k) / D(1,k)$, where $\Delta_{\rm m}(a,k)$ is the matter density perturbation. $P(k)$ is the matter power spectrum today.

The Limber approximation becomes accurate at the percent level for $\ell\gtrsim10$. This condition is satisfied by about 90\% of the total 42 data points that are used in the {\sc iswwll} code. The data are divided into nine galaxy sample bins $j$, i.e., 2MASS0-3, LRG0-1, QSO0-1, and NVSS. The function $f_j(z)$ relates the matter density to the observed projected galaxy overdensity with $f_j(z)=b_j(z)\Pi_j(z)$ in the absence of magnification bias. $\Pi_j(z)$ is the redshift distribution of the galaxies and the bias factor $b_j(z)$ is assumed independent of scale (cf.~\cite{smith:09}), but dependent on redshift. The code determines $f_j(z)$, among other things, from fitting auto power spectra and cross power spectra between the samples.
We refer to Appendix~\ref{sec:iswwll} for more details about the data and the accuracy of the Limber approximation.

\begin{figure*}
\resizebox{\hsize}{!}{\includegraphics{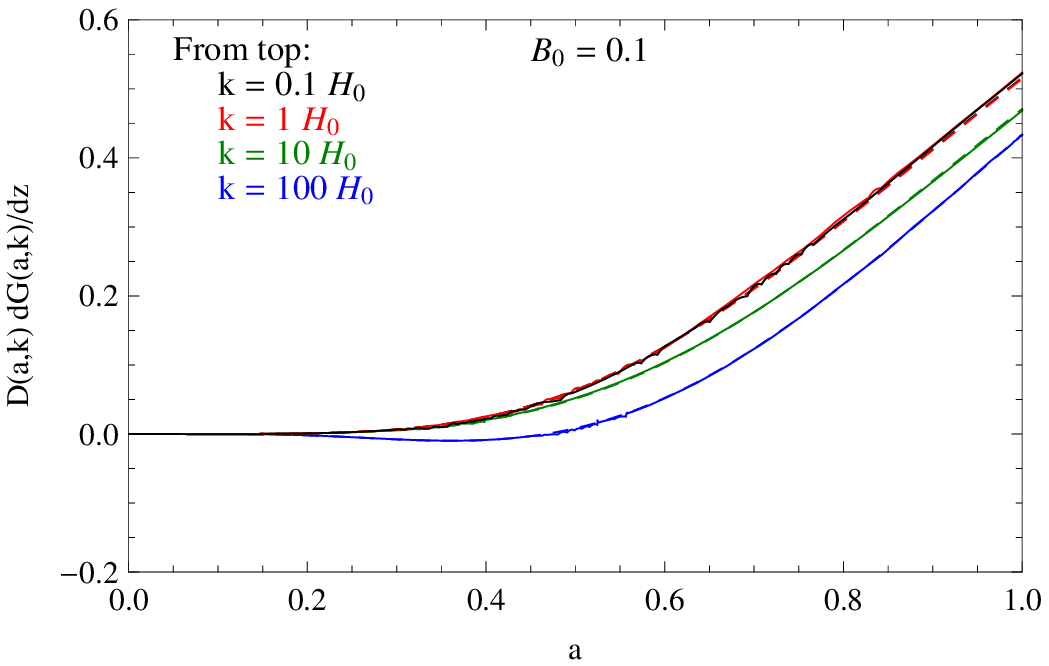}\includegraphics{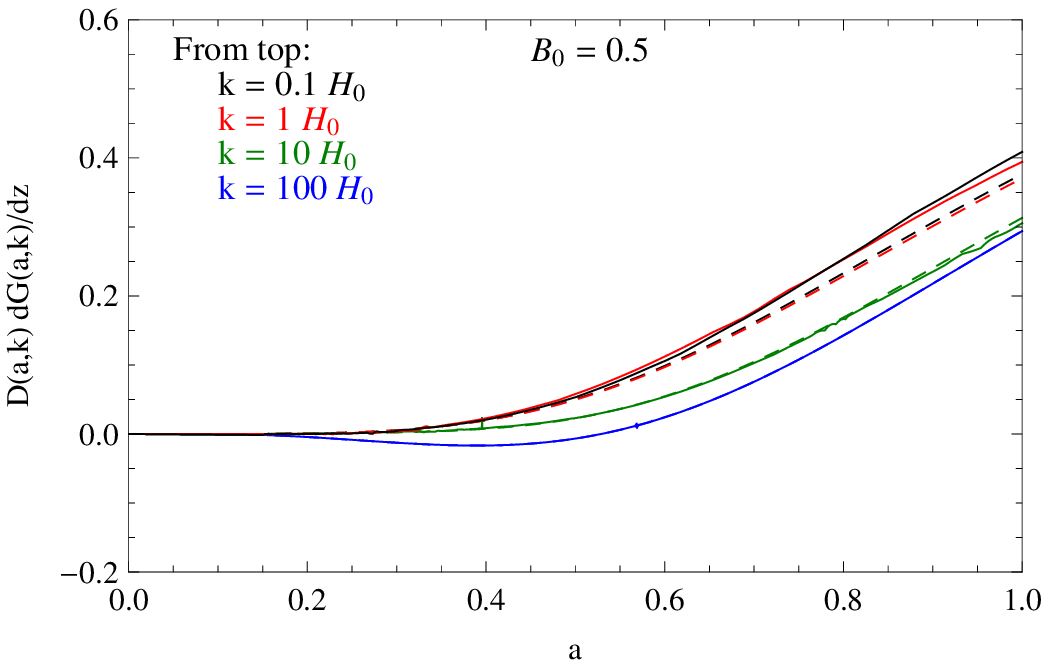}}
\resizebox{\hsize}{!}{\includegraphics{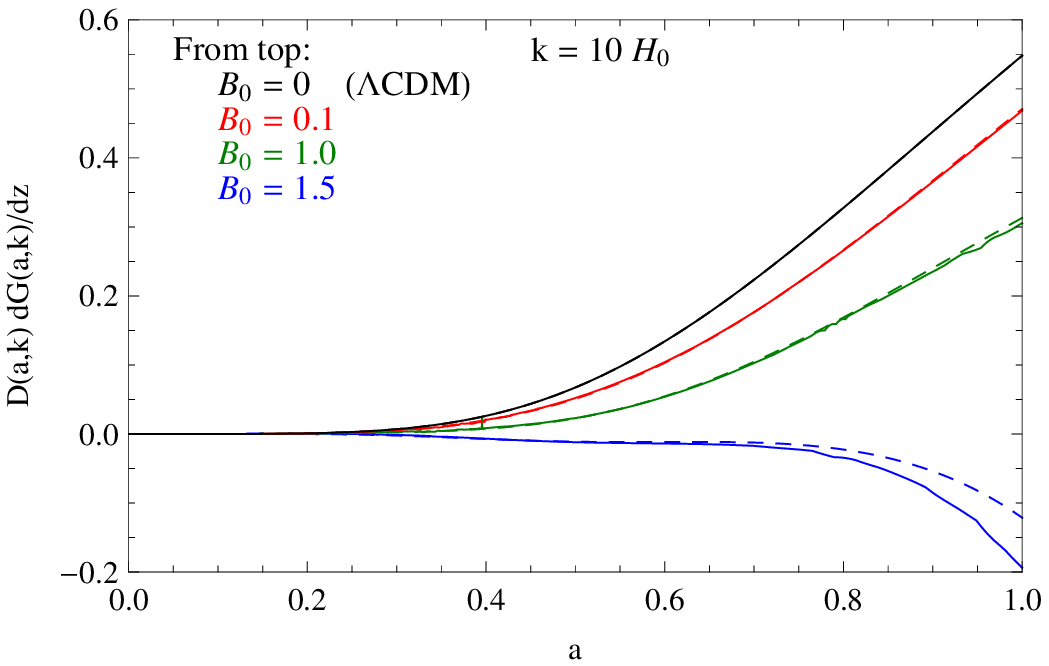}\includegraphics{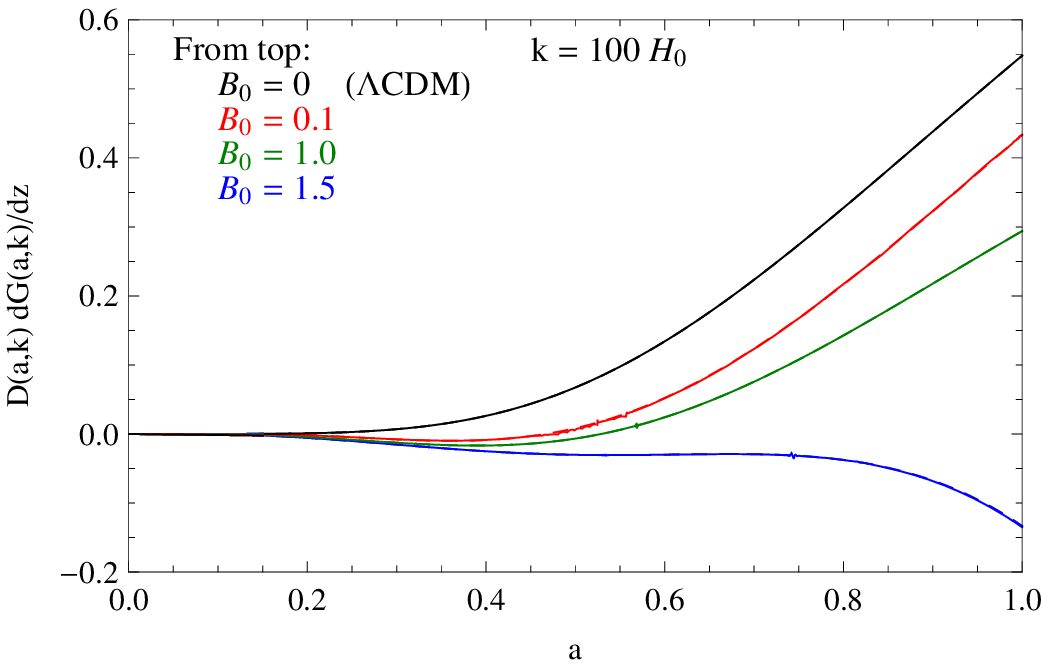}}
\caption{Product of the linear density growth rate $D$ and the derivative of the linear potential growth rate $G$ with respect to redshift $z$. Solid lines are derived from linear perturbation theory, dashed lines are obtained through the approximation described in Sec.~\ref{gISW}, using $h=0.73$ and $\Omega_{\rm m} = 0.24$. Note that $H_0 \simeq 3\times10^{-4}h~\textrm{Mpc}^{-1}$.}
\label{fig:QSapprox}
\end{figure*}

Scalar linear perturbations of the Friedmann metric are presented here in longitudinal gauge, i.e.,
\begin{equation}
ds^2 = -(1+2\Psi) dt^2 + a^2 (1+2\Phi) dx^2,
\end{equation}
where $dx^2$ is the unperturbed spatial line element with curvature $K=0$. We define $\Phi_- \equiv (\Phi - \Psi)/2$.
In the quasistatic regime, we infer
\begin{eqnarray}
G(a,k) & = & \frac{\Phi_-(a,k)}{\Phi_-(a_i,k)} \frac{\Delta_{\rm m}(a_i,k)}{\Delta_{\rm m}(1,k)} \frac{1}{a_i} \nonumber\\
& \simeq & \frac{1}{1+f_R} \frac{D(a,k)}{D(1,k)} \frac{1}{a}
\label{eq:Gak}
\end{eqnarray}
from the modified Poisson equation,
\begin{equation}
k^2\Phi_-(a,k) = \frac{1}{2} \frac{\kappa^2}{1+f_R} \frac{H_0^2 \Omega_{\rm m}}{a} \Delta_{\rm m}(a,k),
\end{equation}
requiring that the initial conditions lie well within the high-curvature regime, where general relativity holds. This implies $\Phi_-(a_i,k) \simeq \Phi_-(a_i)$, $D(a_i,k) \simeq D(a_i)$, and $f_R(a_i) \simeq 0$. We solve the ordinary differential equation
\begin{equation}
\Delta_{\rm m}'' + \left( 2+\frac{H'}{H} \right) \Delta_{\rm m}' - \frac{3}{2} \frac{ 1-g(a,k) }{1+f_R} \frac{H_0^2 \Omega_{\rm m}}{H^2 a^3} \Delta_{\rm m} = 0
\label{eq:Delta_ODE}
\end{equation}
for the linear matter density perturbation $\Delta_{\rm m}(a,k)$, where $g(a,k)$ is the metric ratio in PPF formalism (see Appendix~\ref{sec:PPF}). Note that in the limit $B_0 \rightarrow 0$, we have $g \rightarrow 0$. Therefore, in this limit, Eq.~(\ref{eq:Delta_ODE}) recovers the quasistatic ordinary differential equation for the matter overdensity in $\Lambda$CDM. We solve Eq.~(\ref{eq:Delta_ODE}) with initial conditions at $a_i \ll 1$, in a regime where general relativity is expected to hold, i.e., $\Delta_{\rm m}'(a_i,k)=\Delta_{\rm m}(a_i,k)$ with a normalization set by the initial power spectrum.
At the scales that are relevant for the gISW cross correlations, the product $D\:dG/dz$ used in Eq.~(\ref{eq:gISW}) is accurately described through solving Eq.~(\ref{eq:Delta_ODE}) for $\Delta_{\rm m}$ and using the approximated $G(a,k)$ of Eq.~(\ref{eq:Gak}). We show this by comparing the approximated product to its counterpart from a full derivation within linear perturbation theory (see Fig.~\ref{fig:QSapprox}). We take the relations that exactly describe the scalar linear perturbation theory in $f(R)$ gravity for a matter-only universe from Ref.~\cite{song:06}.

\subsubsection{Weak gravitational lensing and galaxy flows}

The relationship of weak gravitational lensing around galaxies to their large-scale velocities has been proposed as a smoking gun of gravity~\cite{zhang:07}. The advantage of such a probe lies in its insensitivity to galaxy bias and initial matter fluctuations. The expectation value of the ratio of galaxy-galaxy to galaxy-velocity cross correlations of the same galaxies yields an estimator~\cite{zhang:07}
\begin{equation}
E_G = \frac{\Omega_{\rm m}}{(1+f_R)\beta},
\label{eq:eg}
\end{equation}
where $\beta\equiv d \ln \Delta_{\rm m} / d \ln a$. Recently this quantity has been measured analyzing $70\,205$ luminous red galaxies~\cite{eisenstein:01} from the SDSS~\cite{york:00}, yielding $E_G=0.392\pm0.065$~\cite{reyes:10} by averaging over scales $R=(10-50)h^{-1}~\textrm{Mpc}$. Note that the $E_G$ measurement is based on spectroscopic LRG samples, whereas the gISW analysis uses photometric LRG samples that do not overlap in redshift. Furthermore, the error on $E_G$ is dominated by uncertainties in lensing and redshift space distortions and most of the signal comes from small scales around $10h^{-1}~\textrm{Mpc}$. The gISW signal is dominated by large scales and most of the error is caused by sampling variance and shot noise of galaxies. Therefore, we can safely neglect correlations between the $E_G$ and gISW data sets.

\begin{figure}[!b]
%\resizebox{\hsize}{!}{\includegraphics{EG1.eps}\includegraphics{EG2.eps}}
\resizebox{\hsize}{!}{\includegraphics{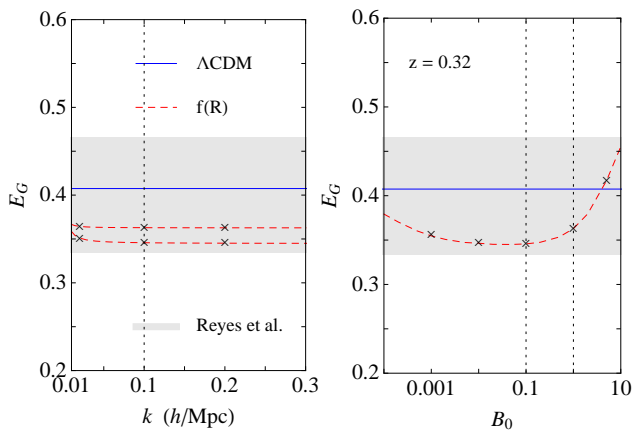}}
\caption{Best-fit $\Lambda$CDM ($B_0=0$) $E_G$ prediction from linear theory (blue solid line). Left panel: adding nonzero values for the Compton wavelength parameter (red dashed lines): $B_0=1$ (top) and $B_0=0.1$ (bottom). Right panel: $E_G$ at different values of $B_0$, evaluated at $k=0.1h~\textrm{Mpc}^{-1}$.}
\label{fig:EG}
\end{figure}

We calculate $\Delta_{\rm m}$ from solving Eq.~(\ref{eq:Delta_ODE}), which yields a good approximation to $E_G$ for the scales and Compton wavelength parameters of interest. We illustrate predictions for $E_G$ in Fig.~\ref{fig:EG}. 
The red-dashed lines show the approximated values and crosses indicate check points derived using full linear perturbation theory.
At the redshift of the measurement, $z=0.32$, and in the linear regime of $f(R)$ gravity, Eq.~(\ref{eq:eg}) is only weakly dependent on scale and shows no $k$-dependence at all for $B_0=0$. Therefore, we only need to evaluate $E_G$ at a representative scale, which we choose here as $k=0.1h~\textrm{Mpc}^{-1}$. We can then compare this value to the mean $E_G$ observation  from~\cite{reyes:10}. At small values of the Compton wavelength parameter ($B_0\lesssim0.01$), a $k$-dependence shows up for $k\lesssim0.1h~\textrm{Mpc}^{-1}$. However, the effect is small compared to the error in $E_G$ and is subdominant to the uncertainty in $\Omega_{\rm m}$.

Fig.~\ref{fig:EG} demonstrates that the $E_G$ probe has the potential to discriminate between $\Lambda$CDM and $f(R)$ gravity even at low values of $B_0$.
%Once the data improves, this will, however, also require correct modeling of the chameleon effect, which, at this stage of the measurement, can safely be neglected.

\subsubsection{Cluster abundance} \label{sec:clusterabundance}

Tighter constraints on modified gravity theories and in particular on $f(R)$ gravity than from the linear theory
probes can be achieved by testing the weakly to fully nonlinear
scales (cf.~\cite{schmidt:09, rapetti:09, daniel:10a}). We use the abundance of clusters to set our strongest boundary
on the possible $B_0$ values. We employ a preliminary version of the likelihood code
of~\cite{seljak:09}, which utilizes constraints from the most massive
halos ($M > 10^{13}~M_{\odot}/h$) inferred from SDSS
data. The galaxy-galaxy lensing signal from clusters and groups from
the MaxBCG catalog~\cite{koester:07} is measured in three cumulative mass bins
corresponding to the nominal number densities of groups of $2.5\times
10^{-7}~({\rm Mpc}/h)^{-3}$, $2\times 10^{-6}~({\rm Mpc}/h)^{-3}$,
and $1.8\times 10^{-5}~({\rm Mpc}/h)^{-3}$ and in two redshift bins
of $z=0.18$ and $z=0.25$. This signal is compared to the theoretical
predictions based on the mass function calibrated with $N$-body
simulations, correctly taking into account mass-observable scatter,
calibration uncertainties and covariances. We refer to Appendix~\ref{sec:cluster-abundance-fr}
for more details.

For comparison, we additionally analyze constraints from applying the
$\sigma_8$ measurement of~\cite{vikhlinin:08}
\begin{equation}
\sigma_8 \left( \frac{\Omega_{\rm m}}{0.25} \right)^{0.47} = 0.813 \pm 0.013 \;(\textrm{stat}) \pm 0.024 \;(\textrm{sys})
\label{eq:vikhlinin}
\end{equation}
inferred from \textit{Chandra} observations of X-ray galaxy cluster
samples detected in the \textit{ROSAT} All-Sky Survey by normalizing
the mass function at low redshifts. The halos in the sample have masses $M > 10^{12}~M_{\odot}/h$. This data was
used by~\cite{schmidt:09} to put a constraint of
$|f_{R0}|< \left( 1.3^{+1.7}_{-0.6} \right) \times10^{-4}$ ($B_0\lesssim10^{-3}$) (95\% C.L.) on the
Hu-Sawicki~\cite{hu:07a} $f(R)$ gravity model. The range indicates a $\mp9\%$ mass calibration error corresponding to the systematic
error in Eq.~(\ref{eq:vikhlinin}). Note that in the likelihood analysis, we add the systematic error in quadrature. The constraint is obtained
from using modified forces in the spherical collapse calculations. Using standard spherical collapse instead reduces this number to
$|f_{R0}|< 0.4 \times10^{-4}$ (95\% C.L.)~\cite{schmidt:09}. An estimate for the range induced from the mass calibration error can be obtained through scaling from the former result, i.e., $|f_{R0}| \lesssim \left( 0.4^{+0.5}_{-0.2} \right) \times10^{-4}$. This corresponds to an upper bound of $|f_{R0}| \lesssim 1.4\times10^{-4}$ at the 95\% C.L. of both the statistical and systematic error.

\subsection{Constraints} \label{sec:results}

Our basic cosmological parameter set we use in the MCMC analysis is $P=\left\{ \Omega_bh^2, \Omega_ch^2, \theta, \tau, n_s, \ln[10^{10} A_s] \right\}$. On this set we implement the following flat priors: $\Omega_bh^2\in(0.01,0.1)$, $\Omega_ch^2\in(0.045,0.99)$, $\theta\in(0.5,10)$, $\tau\in(0.01,0.8)$, $n_s\in(0.5,1.5)$, and $\ln[10^{10} A_s]\in(2.7,4)$. For $f(R)$ gravity models, where $P \rightarrow P \cup \{B_0\}$, we set a flat prior on the free Compton wavelength parameter of $B_0\in(0,10)$.

We begin with the analysis of $\Lambda$CDM in Tables~\ref{tab:res_flat_lcdm_1} and \ref{tab:res_flat_lcdm_2}. We show constraints of using separately gISW, $E_G$, CA, and $E_G$ \& CA data together with the CMB, SHOES, BAO, and Union measurements, as well as when using all of the data jointly. We also quote maximum likelihood parameters and value. Horizontal lines divide the chain parameters from the derived parameters and the best-fit (maximum) likelihood. Notice the improved constraints for $\Lambda$CDM that are obtained by the inclusion of the CA data. This analysis sets the baseline to which adding the $f(R)$ degree of freedom should be measured.

\begin{figure}[!t]
\resizebox{0.9625\hsize}{!}{\includegraphics{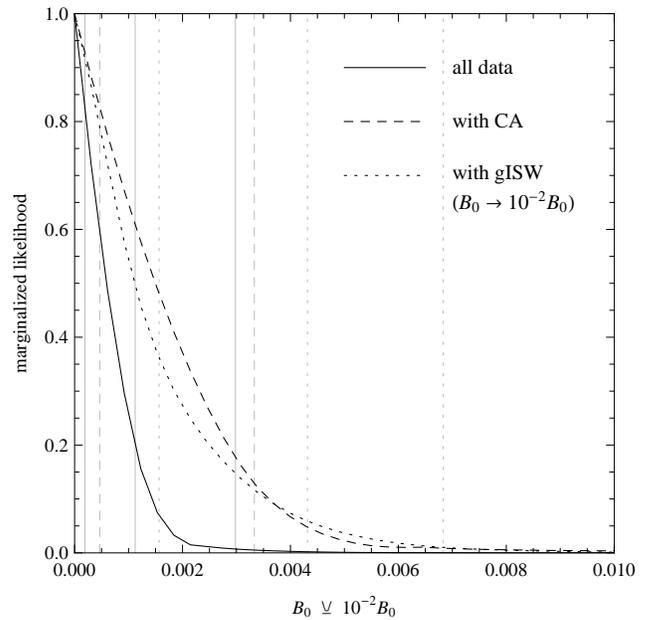}}
\caption{Marginalized likelihood for $B_0$ when using WMAP5, ACBAR, CBI, VSA, Union, SHOES, and BAO in combination with the additional data sets. For gISW, the Compton wavelength parameter is rescaled as $B_0\rightarrow 10^{-2}B_0$, i.e., the constraint is a factor of $10^2$ weaker than illustrated. The horizontal lines indicate 68\%, 95\%, and 99\% confidence levels.}
\label{fig:marg}
\end{figure}

\begin{figure}[!t]
\resizebox{\hsize}{!}{\includegraphics{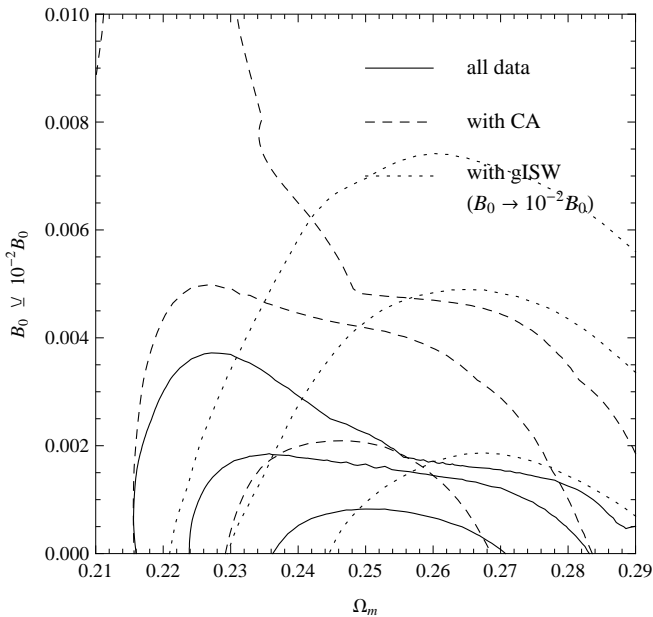}}
\caption{Contours of 2D marginalized 68\%, 95\%, and 99\% confidence boundaries using WMAP5, ACBAR, CBI, VSA, Union, SHOES, and BAO in combination with the additional data sets. For gISW, the Compton wavelength is rescaled as $B_0\rightarrow 10^{-2}B_0$, i.e., the constraint is a factor of $10^2$ weaker than illustrated. gISW cross correlations favor higher values of $\Omega_{\rm m}$ and break the degeneracy between $\Omega_{\rm m}$ and $B_0$ seen in the CA data.}
\label{fig:cont}
\end{figure}

In this paper, we study flat metric $f(R)$ gravity models that reproduce the $\Lambda$CDM expansion history. Therefore, the SHOES, BAO, and Union measurements only fix the background and do not distinguish standard from modified gravity. However, they contribute to the breaking of degeneracies that show up in other data sets and help tightening the constraints. When introducing the CMB probes, some additional information becomes available. In fact, we observe that $f(R)$ gravity yields a slightly better fit than flat $\Lambda$CDM, which can be attributed to the lowering of the temperature anisotropy power spectrum at small $\ell$ (see Fig.~\ref{fig:CMB}), but the improvement in the fit is not at a significant level (see  Table~\ref{tab:res_flat_fR_1}).
In contrast to flat $\Lambda$CDM, where the inclusion of the gISW data does not yield noticeable improvement on the parameter constraints~\cite{ho:08}, in the case of $f(R)$ gravity, an order of magnitude improvement on the $B_0$ constraint is achieved, i.e., $B_0<0.42$ at the 95\% C.L.
This constraint is in perfect agreement with the independent result of~\cite{giannantonio:09}, who found an upper bound on the Compton wavelength parameter of $B_0<0.4$ at the 95\% C.L., using gISW cross correlations data from~\cite{giannantonio:08}.
Including the cluster abundance data instead yields another two orders of magnitude improvement over the gISW constraint, i.e., $B_0<3.33\times10^{-3}$. Here and throughout this section, we quote constraints at the 95\% C.L. If we add either the gISW or $E_G$ data, this bound tightens by a factor of 1.9 or 2.2, respectively. The joint constraint on the Compton wavelength parameters of $B_0<1.12\times10^{-3}$ (a factor of 3.0) (see  Table~\ref{tab:res_flat_fR_2}) is our main result and is inferred from combining all of the data sets. We therefore found that the gISW and $E_G$ probes are good complementary tests to cluster abundance. In Fig.~\ref{fig:marg} and Fig.~\ref{fig:cont}, we plot the marginalized likelihood for $B_0$ and the 2D-marginalized contours for $B_0$ and $\Omega_{\rm m}$, respectively, with different combinations of the data sets, illustrating this point. It applies particularly to $E_G$, where bounds on $B_0$ become looser in the absence of data from cluster abundance. This is what one expects from the trend seen in Fig.~\ref{fig:EG}, i.e., for $B_0\gtrsim0.1$ the $E_G$ prediction approaches its $\Lambda$CDM value and eventually overshoots it.

If we additionally include CCCP to our joint set of data, we obtain a constraint of $B_0<0.96\times10^{-3}$ or equivalently $|f_{R0}|<1.65\times10^{-4}$. If we take CCCP with neither gISW, $E_G$, nor CA the constraint becomes $B_0<1.83\times10^{-3}$ or equivalently $|f_{R0}|<3.12\times10^{-4}$. The constraint is a factor of 2.2 times weaker than what we estimated from the result of~\cite{schmidt:09} when adding the systematic error from mass calibration in the case of standard spherical collapse (see Sec.~\ref{sec:clusterabundance}).
%, which is an order of magnitude less than the result of Ref.~\cite{schmidt:09} when the spherical collapse parameters are not modified. The discrepancy is most likely due to using Eq.~(\ref{eq:vikhlinin}) rather then the full likelihood function of~\cite{vikhlinin:08}, but it could also originate from doing a full MCMC analysis over all cosmological parameters rather than marginalizing over several parameters, from a mass calibration error, or a combination of all three. We stress however that our main result, although a little looser, is in good agreement with the main result of Ref.~\cite{schmidt:09}.
Note that we have used $\sigma_8$ predicted directly by $f(R)$ gravity in Eq.~(\ref{eq:vikhlinin}) rather than constraining the rescaled normalization $\sigma_8^{\rm eff}$ in $\Lambda$CDM that matches the halo mass function in $f(R)$ gravity at the pivot mass $M_{\rm eff} = 3.667\times10^{14}~\textrm{M}_{\odot}/h$ ($M=M_{500}$) (cf.~\cite{schmidt:09}). In the case of standard spherical collapse, this leads to an overestimation of $f(R)$ gravity effects on $\sigma_8$ of $\sim2\%$ for $B_0=10^{-3}$. The error of the CCCP measurement in Eq.~(\ref{eq:vikhlinin}) is $\sim\pm3\%$.
Also note that the CA data infer a preliminary constraint of~\cite{seljak:09}
\begin{equation}
\sigma_8 \left( \frac{\Omega_{\rm m}}{0.25} \right)^{0.40} = 0.844 \pm 0.036,
\end{equation}
which is a slightly larger value than in Eq.~(\ref{eq:vikhlinin}) and therefore admits larger values of $B_0$, i.e., using CCCP instead of CA (without gISW and $E_G$) tightens the 95\% C.L. boundary on $B_0$ by a factor of 1.8. It is also important to keep in mind that the constraint of~\cite{schmidt:09} is inferred for the Hu-Sawicki model that exhibits an enhanced linear growth with respect to the designer model studied here (see, e.g.,~\cite{ferraro:10}).

Finally, note that since $\Lambda$CDM is reproduced in the limit $B_0\rightarrow0$, the slightly poorer fits of $f(R)$ gravity with respect to $\Lambda$CDM (see Tables~\ref{tab:res_flat_fR_1} and \ref{tab:res_flat_fR_2}) have to be attributed to sampling errors in the chains.

%\begin{turnpage}

\begin{table*}[ht]
%\resizebox{\hsize}{!}{
\centering
\begin{tabular}{|l|r@{}|c|c|r@{}|c|c|r@{}|c|c|}
\hline
%\cline{1-1} \cline{3-4} \cline{6-7}
Parameters & & \multicolumn{2}{|c|}{$\Lambda$CDM} & & \multicolumn{2}{|c|}{$\Lambda$CDM (with gISW)} & & \multicolumn{2}{|c|}{$\Lambda$CDM (with $E_G$)}  \\
\cline{1-1} \cline{3-4} \cline{6-7} \cline{9-10}
$100\Omega_b h^2$  & & $2.235\pm0.053$   & 2.232  & & $2.235\pm0.054$   & 2.237  & & $2.235\pm0.054$   & 2.228   \\
$\Omega_c h^2$     & & $0.1128\pm0.0036$ & 0.1137 & & $0.1124\pm0.0036$ & 0.1125 & & $0.1127\pm0.0037$ & 0.1131  \\
$\theta$           & & $1.0407\pm0.0026$ & 1.0402 & & $1.0406\pm0.0027$ & 1.0406 & & $1.0406\pm0.0027$ & 1.0409  \\
$\tau$             & & $0.082\pm0.015$   & 0.079  & & $0.084\pm0.016$   & 0.085  & & $0.082\pm0.016$   & 0.081   \\
$n_s$              & & $0.957\pm0.012$   & 0.955  & & $0.958\pm0.012$   & 0.956  & & $0.957\pm0.012$   & 0.957   \\
$\ln[10^{10}A_s]$  & & $3.202\pm0.038$   & 3.204  & & $3.202\pm0.038$   & 3.210  & & $3.202\pm0.039$   & 3.202   \\
\cline{3-4} \cline{6-7} \cline{9-10}
$\Omega_m$         & & $0.273\pm0.016$   & 0.279  & & $0.271\pm0.016$   & 0.271  & & $0.272\pm0.016$   & 0.273   \\
$H_0$              & & $70.4\pm1.4$      & 69.9   & & $70.6\pm1.3$      & 70.5   & & $70.5\pm1.3$      & 70.3    \\
\cline{1-1} \cline{3-4} \cline{6-7} \cline{9-10}
$-2\ln L$ & & \multicolumn{2}{|c|}{3027.812} & & \multicolumn{2}{|c|}{3061.630} & & \multicolumn{2}{|c|}{3027.826} \\
\cline{1-1} \cline{3-4} \cline{6-7} \cline{9-10}
\end{tabular}
%}
\caption{Means, standard deviations (left subdivision of columns) and best-fit values (right subdivision of columns) with likelihood for the flat $\Lambda$CDM model using data from WMAP, ACBAR, CBI, VSA, Union, BAO, and SHOES (left column). Including one the gISW (middle column) and $E_G$ data sets (right column).}
\label{tab:res_flat_lcdm_1}
\end{table*}

\begin{table*}[ht]
%\resizebox{\hsize}{!}{
\centering
\begin{tabular}{|l|r@{}|c|c|r@{}|c|c|r@{}|c|c|}
\hline
%\cline{1-1} \cline{3-4} \cline{6-7}
Parameters & & \multicolumn{2}{|c|}{$\Lambda$CDM (with CA)}  & & \multicolumn{2}{|c|}{$\Lambda$CDM (with $E_G$\&CA)}  & & \multicolumn{2}{|c|}{$\Lambda$CDM (all)}  \\
\cline{1-1} \cline{3-4} \cline{6-7} \cline{9-10}
$100\Omega_b h^2$  & & $2.228\pm0.051$   & 2.233  & & $2.229\pm0.053$   & 2.229  & & $2.231\pm0.053$   & 2.239 \\
$\Omega_c h^2$     & & $0.1107\pm0.0019$ & 0.1112 & & $0.1107\pm0.0020$ & 0.1112 & & $0.1107\pm0.020$  & 0.1108 \\
$\theta$           & & $1.0401\pm0.0025$ & 1.0400 & & $1.0402\pm0.0025$ & 1.0406 & & $1.0403\pm0.0026$ & 1.0408 \\
$\tau$             & & $0.081\pm0.015$   & 0.080  & & $0.081\pm0.015$   & 0.078  & & $0.082\pm0.015$   & 0.078 \\
$n_s$              & & $0.956\pm0.012$   & 0.957  & & $0.956\pm0.012$   & 0.956  & & $0.956\pm0.012$   & 0.957 \\
$\ln[10^{10}A_s]$  & & $3.195\pm0.035$   & 3.191  & & $3.195\pm0.036$   & 3.193  & & $3.196\pm0.036$   & 3.185 \\
\cline{3-4} \cline{6-7} \cline{9-10}
$\Omega_m$         & & $0.2642\pm0.0099$ & 0.2664 & & $0.2638\pm0.0098$ & 0.2654 & & $0.2634\pm0.0098$ & 0.2622 \\
$H_0$              & & $71.0\pm1.1$       & 70.8   & & $71.0\pm1.1$       & 70.9   & & $71.1\pm1.1$      & 71.3 \\
\cline{1-1} \cline{3-4} \cline{6-7} \cline{9-10}
$-2\ln L$ & & \multicolumn{2}{|c|}{3032.706} & & \multicolumn{2}{|c|}{3032.746} & & \multicolumn{2}{|c|}{3066.240} \\
\cline{1-1} \cline{3-4} \cline{6-7} \cline{9-10}
\end{tabular}
%}
\caption{Same as Table~\ref{tab:res_flat_lcdm_1}, but including the CA (left column), both $E_G$ and CA (middle column), and all (right column) additional data sets.}
\label{tab:res_flat_lcdm_2}
\end{table*}

\begin{table*}[ht]
%\resizebox{\hsize}{!}{
\centering
\begin{tabular}{|l|r@{}|c|c|r@{}|c|c|r@{}|c|c|}
\hline
%\cline{1-1} \cline{3-4} \cline{6-7}
Parameters & & \multicolumn{2}{|c|}{$f(R)$} & & \multicolumn{2}{|c|}{$f(R)$ (with gISW)} & & \multicolumn{2}{|c|}{$f(R)$ (with $E_G$)} \\
\cline{1-1} \cline{3-4} \cline{6-7} \cline{9-10}
$100\Omega_b h^2$  & & $2.223\pm0.053$   & 2.206  & & $2.225\pm0.054$   & 2.253  & & $2.224\pm0.054$   & 2.206  \\
$\Omega_c h^2$     & & $0.1123\pm0.0036$ & 0.1109 & & $0.1117\pm0.0036$ & 0.1133 & & $0.1125\pm0.0036$ & 0.1131 \\
$\theta$           & & $1.0403\pm0.0027$ & 1.0392 & & $1.0403\pm0.0027$ & 1.0416 & & $1.0403\pm0.0027$ & 1.0394 \\
$\tau$             & & $0.083\pm0.016$   & 0.082  & & $0.084\pm0.016$   & 0.090  & & $0.083\pm0.016$   & 0.083  \\
$n_s$              & & $0.954\pm0.012$   & 0.950  & & $0.954\pm0.012$   & 0.965  & & $0.954\pm0.013$   & 0.952  \\
$\ln[10^{10}A_s]$  & & $3.212\pm0.040$   & 3.215  & & $3.209\pm0.039$   & 3.200  & & $3.213\pm0.039$   & 3.221  \\
$100 B_0$          & & $<315$            & 28     & & $<43.2$           & 0.0    & & $<319$            & 30     \\
\cline{3-4} \cline{6-7} \cline{9-10}
$\Omega_m$         & & $0.272\pm0.016$   & 0.268  & & $0.269\pm0.016$   & 0.272  & & $0.273\pm0.016$   & 0.279  \\
$H_0$              & & $70.4\pm1.4$      & 70.4   & & $70.7\pm1.3$      & 70.7   & & $70.3\pm1.3$      & 69.6   \\
$10^3|f_{R0}|$     & & $<350$            & 46     & & $<69.4$           & 0.0    & & $<353$            & 51     \\
\cline{1-1} \cline{3-4} \cline{6-7} \cline{9-10}
$-2\Delta\ln L$ & & \multicolumn{2}{|c|}{-1.104} & & \multicolumn{2}{|c|}{1.506} & & \multicolumn{2}{|c|}{-0.696} \\
\cline{1-1} \cline{3-4} \cline{6-7} \cline{9-10}
\end{tabular}
%}
\caption{Same as Table~\ref{tab:res_flat_lcdm_1}, but for $f(R)$ gravity. $-2\Delta\ln L$ is quoted with respect to the corresponding maximum likelihood flat $\Lambda$CDM model. Limits on $B_0$ and $|f_{R0}|$ indicate the one-sided 1D marginalized upper 95\% C.L.  Note that as $B_0\rightarrow0$ reproduces $\Lambda$CDM predictions, the slightly poorer fits of $f(R)$ gravity should be attributed to sampling error in the MCMC runs.}
\label{tab:res_flat_fR_1}
\end{table*}

\begin{table*}[ht]
%\resizebox{\hsize}{!}{
\centering
\begin{tabular}{|l|r@{}|c|c|r@{}|c|c|r@{}|c|c|}
\hline
%\cline{1-1} \cline{3-4} \cline{6-7}
Parameters & & \multicolumn{2}{|c|}{$f(R)$ (with CA)}  & & \multicolumn{2}{|c|}{$f(R)$ (with $E_G$\&CA)}  & & \multicolumn{2}{|c|}{$f(R)$ (all)} \\
\cline{1-1} \cline{3-4} \cline{6-7} \cline{9-10}
$100\Omega_b h^2$  & & $2.209\pm0.054$   & 2.204  & & $2.213\pm0.054$   & 2.235  & & $2.216\pm0.054$   & 2.210 \\
$\Omega_c h^2$     & & $0.1064\pm0.0032$ & 0.1112 & & $0.1073\pm0.0029$ & 0.1108 & & $0.1076\pm0.0028$ & 0.1104 \\
$\theta$           & & $1.0390\pm0.0027$ & 1.0398 & & $1.0392\pm0.0027$ & 1.0413 & & $1.0394\pm0.0027$ & 1.0398 \\
$\tau$             & & $0.077\pm0.016$   & 0.080  & & $0.077\pm0.015$   & 0.084  & & $0.079\pm0.015$   & 0.075 \\
$n_s$              & & $0.953\pm0.012$   & 0.951  & & $0.954\pm0.012$   & 0.956  & & $0.954\pm0.012$   & 0.951 \\
$\ln[10^{10}A_s]$  & & $3.175\pm0.0038$  & 3.209  & & $3.179\pm0.037$   & 3.203  & & $3.182\pm0.0037$  & 3.193 \\
$100 B_0$          & & $<0.333$          & 0.000  & & $<0.152$          & 0.000  & & $<0.112$          & 0.001 \\
\cline{3-4} \cline{6-7} \cline{9-10}
$\Omega_m$         & & $0.247\pm0.014$   & 0.268  & & $0.251\pm0.012$   & 0.261  & & $0.252\pm0.012$   & 0.264 \\
$H_0$              & & $72.2\pm1.4$      & 70.5   & & $71.9\pm1.3$      & 71.4   & & $71.9\pm1.2$      & 70.8 \\
$10^3|f_{R0}|$     & & $<0.484$          & 0.001  & & $<0.263$          & 0.000  & & $<0.194$          & 0.002 \\
\cline{1-1} \cline{3-4} \cline{6-7} \cline{9-10}
$-2\Delta\ln L$ & & \multicolumn{2}{|c|}{0.802} & & \multicolumn{2}{|c|}{0.264} & & \multicolumn{2}{|c|}{0.926} \\
\cline{1-1} \cline{3-4} \cline{6-7} \cline{9-10}
\end{tabular}
%}
\caption{Same as Table~\ref{tab:res_flat_lcdm_2}, but for $f(R)$ gravity. See also Table~\ref{tab:res_flat_fR_1}.}
\label{tab:res_flat_fR_2}
\end{table*}

%\end{turnpage}

\section{Discussion} \label{sec:discussion}

We have performed a MCMC analysis on metric $f(R)$ gravity models that exactly reproduce the $\Lambda$CDM expansion history. In addition to geometrical probes from supernovae, BAO distance, and Hubble constant measurements, which were used to fix the background, we utilized all of the CMB data, including the lowest multipoles, its correlation with galaxies, the comparison of weak gravitational lensing to large-scale velocities, and the abundance of clusters.

We report a constraint on the Compton wavelength parameter of $B_0<1.1\times10^{-3}$ at the 95\% C.L. from using all of the measurements. This result is substantially driven by data from cluster abundance. However as the data improve, the limits will saturate due to the chameleon effect in massive haloes. gISW measures in combination with the CMB, supernovae, BAO distance, and Hubble constant probes, yield a constraint of $B_0<0.42$ (95\% C.L.), which is an order of magnitude improvement over using the CMB alone as probe of the growth of large-scale structure in combination with the geometrical measures. This highlights the power of gISW measurements as a linear theory probe to constrain infrared modifications of gravity.

The $E_G$ measurement of the relationship between weak gravitational lensing and galaxy flows does not improve bounds on $f(R)$ gravity on its own. However, when used as a complementary probe to cluster abundance, it contributes substantially to our constraints. This can be attributed to the slow convergence of its prediction toward $\Lambda$CDM when $B_0\rightarrow0$. It is likely that with improved data, the $E_G$ probe will become an important discriminator for gravity models.

\section*{Acknowledgments}

We would like to thank Fabian Schmidt, Wenjuan Fang, and Tristan Smith
for useful discussions and Doug Potter for technical
support. We are very grateful to Rachel Mandelbaum for providing preliminary results for the galaxy-galaxy lensing signal from clusters and groups from the MaxBCG catalog. Computational resources were provided on the zBox2, zBox3,
and Schr\"{o}dinger supercomputers at the University of
Z\"{u}rich. This work was supported by the
Swiss National Foundation under Contract No.~2000\_124835/1, WCU
Grant No.~R32-2008-000-10130-0, and the U.S. Department of Energy under
Contract No. DE-AC02-98CH10886. WH was supported by the KICP under
NSF contract PHY-0114422, DOE contract DE-FG02-90ER-40560 and the Packard Foundation.

\appendix

\section{PPF linear theory} \label{sec:PPF}

Given the expansion history, the PPF framework~\cite{hu:07b, hu:08} is defined by three functions and one parameter. From these quantities, the dynamics are determined by conservation of energy and momentum and the Bianchi identities. The defining quantities are $g(a,k)$, which quantifies the effective anisotropic stress of the modifications and distinguishes the two gravitational potentials, $f_{\zeta}(a)$, which describes the relationship between the matter and the metric on superhorizon scales, and $f_G(a)$, which defines it in the linearized Newtonian regime. The additional parameter $c_{\Gamma}$ is the transition scale between the superhorizon and Newtonian behaviors.
For $f(R)$ gravity the PPF expressions were developed in~\cite{hu:07b}. For completeness, we shall review it here.

At superhorizon scales, the metric ratio
\begin{equation}
g_{\rm SH}(\ln a) = \frac{\Phi+\Psi}{\Phi-\Psi}
\end{equation}
is obtained from solving the differential equation
\begin{eqnarray}
\Phi'' + \left( 1 - \frac{H''}{H'} + \frac{B'}{1-B} + B\frac{H'}{H} \right) \Phi' & & \nonumber\\
+ \left( \frac{H'}{H} - \frac{H''}{H'} + \frac{B'}{1-B} \right) \Phi & = & 0,
\end{eqnarray}
where $k/aH \rightarrow 0$, and using the relation
\begin{equation}
\Psi = \frac{-\Phi-B\Phi'}{1-B},
\end{equation}
where $k/aH \rightarrow 0$~\cite{hu:07b}. This follows from conservation of curvature fluctuation $(\zeta'=0)$ and momentum, considering the superhorizon anisotropy relation $\Phi+\Psi=BH'q$~\cite{song:06}, where $q$ is the momentum fluctuation.

In the quasistatic regime, we have $g_{\rm QS}=-1/3$ and at intermediate scales
\begin{equation}
g(a, k) = \frac{g_{\rm SH}+g_{\rm QS}(c_gk/aH)^{n_g}}{1+(c_gk/aH)^{n_g}},
\end{equation}
where $c_g=0.71\sqrt{B}$ and $n_g=2$~\cite{hu:07b}. Further, $f_{\zeta}=-g_{\rm SH}/3$, $f_G=f_R$ and $c_{\Gamma}=1$~\cite{hu:07b}.

We supply the PPF modified {\sc camb} code~\cite{fang:08} with the functions defined above. This gives a good approximation for $B_0\lesssim1$. Above this value, the approximation begins to break down at intermediate scales and low redshifts, as we have tested by comparing the metric ratio predicted by the PPF functions and the exact numerical solution for the ordinary differential equations describing scalar linear perturbation theory in $f(R)$ gravity for a matter-only universe. The expressions defining the $f(R)$ scalar linear perturbation theory can be found in~\cite{song:06}. This deviation shows up in the low multipoles of the CMB for high $B_0$ and partially manifests itself when comparing constraints on $B_0$ derived from the CMB with the results of~\cite{song:07}. However, when including gISW measures or cluster abundance, the viable values of $B_0$ lie well within the regime where the approximation holds.

\section{The {\sc iswwll} code for $f(R)$ gravity} \label{sec:iswwll}

We use the publicly available {\sc iswwll} code~\cite{ho:08, hirata:08} for our analysis. Note that we have turned off weak lensing contributions in the code, focusing only on the gISW constraints. The 42 data points of gISW cross correlations that are used in the likelihood analysis are collected from
the
Two Micron All Sky Survey (2MASS) extended source catalog (XSC)~\cite{jarrett:00, skrutskie:06}, the luminous red galaxies (LRGs) and photometric quasars (QSO) of the SDSS~\cite{adelman:07}, and the National Radio Astronomy Observatory (NRAO) Very Large Array (VLA) Sky Survey (NVSS)~\cite{condon:98}. They are divided into nine galaxy sample bins $j$ (2MASS0-3, LRG0-1, QSO0-1, and NVSS) based on flux (2MASS) or redshift (LRG and QSO). These data points are a selection of multipole bins from all samples, where the selection is based on the avoidance of nonlinearities and systematic effects from dust extinction, galaxy foregrounds, the thermal Sunyaev-Zel'dovich effect, and point source contamination to affect the gISW cross correlations~\cite{ho:08}.

In the remainder of Appendix~\ref{sec:iswwll}, we discuss the validity of the Limber approximation and elucidate the function $f_j(z)$ that carries information about the redshift distribution and bias in the context of $f(R)$ gravity.

\subsection{Limber approximation}

For $f(R)$ gravity the gISW cross correlations are well described within the approximations given in~\textsection\ref{gISW}. The applicability of approximating the product $D\:dG/dz$ in Eq.~(\ref{eq:gISW}) through Eq.~(\ref{eq:Delta_ODE}) and Eq.~(\ref{eq:Gak}) is a direct consequence of applying the Limber approximation. This can easily be seen from Fig.~\ref{fig:QSapprox} considering the substitution $k\rightarrow(\ell+1/2)/\chi(z)$ at the relevant redshifts. The accuracy of the Limber approximation itself, in the case of $\Lambda$CDM, is at the order of 10\% at $\ell=2$ and drops approximately as $\ell^2$ at higher $\ell$ (see, e.g.,~\cite{smith:09, loverde:08, afshordi:04}). Apart from the multipole $\ell$, the error depends also on the width of the redshift distribution, which changes only little with $f(R)$ effects. Given the large errors of the currently available data points at low $\ell$, we conclude that the Limber approximation is applicable and furthermore very useful since it is numerically faster than an exact integration.

\subsection{Redshift distribution and bias}

A further ingredient in the {\sc iswwll} code is the determination of the function $f_j(z)$. In the Markov chain, $f_j(z)$ is recomputed when changing the cosmological parameter values. The methods by which this function is determined differs for each sample, but they are all based on galaxy clustering data.

The 2MASS galaxies are matched with SDSS galaxies in order to identify their redshifts. To obtain the nonlinear power spectrum, the $Q$ model for nonlinearities~\cite{cole:05} is applied. Then, the code computes the galaxy power spectrum and fits it to measurements, thereby determining the bias $b(z)$ and $Q$. Since the required accuracy for the estimation of bias is only at the order of a few tens of percent~\cite{ho:08}, this processing is also applicable to $f(R)$ gravity. The $Q$ model is also adopted for LRG galaxies, where the redshift probability distribution is inferred with methods described in Ref.~\cite{padmanabhan:05}. For QSO, first, a preliminary estimate for the redshift distribution is deduced by locating a region of sky with high spectroscopic completeness, but simultaneously maintaining a large area. Taking into account magnification bias and fitting $b_j(z)\Pi_j(z)$ using the quasar power spectrum and quasar-LRG cross power yields the desired shape of $f_j(z)$. Note that in $f(R)$ gravity the relationship between the metric combination sensitive to gravitational redshifts and lensing $\Phi_-$ and the density perturbations is modified such that the expression of the lensing window function for magnification effects given in Ref.~\cite{ho:08} is rescaled by $(1+f_R)^{-1} \leq (1+f_{R0})^{-1} \lesssim 1.1$ for $B_0\lesssim1$, where $(1+f_R)^{-1} \geq 1$. Since this is a small effect and magnification bias is subdominant to bias and redshift distribution, it is save to neglect it. Due to the rather low accuracy requirement, we can furthermore adopt the $\Lambda$CDM growth factor in the determination of $b_i(z)\Pi_i(z)$. Finally, the effective redshift distribution of NVSS is obtained from cross correlating with the other samples and $f_j(z)$ is fitted with a $\Gamma$ distribution.

\section{Cluster abundance in $f(R)$ gravity}
\label{sec:cluster-abundance-fr}
The likelihood code of~\cite{seljak:09} utilizes the halo mass
function of Tinker \textit{et al.}~\cite{tinker:08} at $z=0.18$, which
is described by the functions
\begin{equation}
\frac{dn}{dM} = f(\sigma) \frac{\bar{\rho}_m}{M} \frac{d \ln \sigma^{-1}}{dM},
\label{eq:tinker}
\end{equation}
where the distribution $f(\sigma)$ is given by
\begin{equation}
f(\sigma) = A \left[ \left( \frac{\sigma}{b} \right)^{-a} + 1 \right] e^{-c/\sigma^2}
\end{equation}
and
\begin{equation}
\sigma^2 = \int P(k) \hat{W}(kR) k^2 dk.
\end{equation}
Here, $\hat{W}$ is the Fourier transform of the real-space top-hat window function of radius $R$.
The free parameters $A$, $a$, $b$, and $c$ at $z=0.18$ are fitted to $\Lambda$CDM simulations in~\cite{tinker:08}.
We define $M$ as the lensing mass, being the more conservative approach than taking it to be the dynamical mass (see~\cite{schmidt:09, schmidt:10}). 
In the large field limit ($B_0\gtrsim10^{-3}$), where no chameleon mechanism takes place, these relations provide a reasonable fit to $f(R)$ gravity simulations as well, using the same values for the parameters as in $\Lambda$CDM, and correspond to using the Sheth-Tormen~\cite{sheth:99} halo mass function with spherical collapse predicted by standard gravity (see~\cite{schmidt:08, schmidt:09}). Another approach is to alter the spherical collapse calculations, taking care of modified forces~\cite{schmidt:08, schmidt:09}. Standard spherical collapse tends to overestimate the $f(R)$ effects, while the opposite occurs for the modified spherical collapse calculations. Given the constraints that can be achieved from the data, we conclude that it is save to assume the large field limit, neglect the chameleon mechanism, and use standard spherical collapse, i.e., using the $\Lambda$CDM values for the free parameters in the Tinker \textit{et al.}~halo mass function. While choosing modified spherical collapse is the more conservative approach, the standard spherical collapse parameters yield a better fit to the simulations (see~\cite{schmidt:08, schmidt:09}). We remind the reader that the 68\% C.L. inferred from cluster abundance should be taken with caution since it approaches the small field limit ($B_0\lesssim10^{-5}$) and intermediate regime, where chameleon effects appear that cannot be described through Eq.~(\ref{eq:tinker}). A better fit for the halo mass function in $f(R)$ gravity is an objective to future work.

\vfill
\bibliographystyle{arxiv_physrev}
\bibliography{fR}

\end{document}